\documentclass[aps,pra,twocolumn,showpacs,notitlepage,superscriptaddress,letterpaper]{revtex4-1}
\usepackage{graphicx}
\usepackage{amsmath}
\usepackage{esint}
\usepackage{verbatim}
\usepackage{color}
\usepackage{SIunits}
\usepackage{hyperref}

\begin{document}
\newcommand{\fig}[1]{Fig.~#1}
\newcommand{\subfig}[2]{Fig.~#1(#2)}
\newcommand{\silife}[0]{27~$\mu$s}
\newcommand{\soilife}[0]{3.5~$\mu$s}
\newcommand{\sicoh}[0]{6.6~$\mu$s}
\newcommand{\soicoh}[0]{2.2~$\mu$s}
\newcommand{\pd}[0]{\color{red}}
\newcommand{\ajk}[0]{\color{blue}}
\newcommand{\ojp}[0]{\color{green}}

\title{Superconducting qubits on silicon substrates for quantum device integration} %Title of paper



\author{Andrew J. Keller}
\author{Paul B. Dieterle}
\author{Michael Fang}
\author{Brett Berger}
\affiliation{Kavli Nanoscience Institute and Thomas J. Watson Laboratory of Applied Physics, California Institute of Technology, Pasadena, CA 91125}
\affiliation{Institute of Quantum Information and Matter, California Institute of Technology, Pasadena, CA 91125}
\author{Johannes M. Fink}
%\affiliation{Institute of Quantum Information and Matter, California Institute of Technology, Pasadena, CA 91125}
\affiliation{Kavli Nanoscience Institute and Thomas J. Watson Laboratory of Applied Physics, California Institute of Technology, Pasadena, CA 91125}
\affiliation{Institute of Quantum Information and Matter, California Institute of Technology, Pasadena, CA 91125}
\affiliation{Institute of Science and Technology Austria, 3400 Klosterneuburg, Austria}
\author{Oskar Painter}
\affiliation{Kavli Nanoscience Institute and Thomas J. Watson Laboratory of Applied Physics, California Institute of Technology, Pasadena, CA 91125}
\affiliation{Institute of Quantum Information and Matter, California Institute of Technology, Pasadena, CA 91125}
\email{opainter@caltech.edu}
\homepage{http://copilot.caltech.edu}
%\thanks{}


\date{\today}

\begin{abstract}
We present the fabrication and characterization of transmon qubits formed from aluminum Josephson junctions on two different silicon-based substrates: (i) high-resistivity silicon (Si) and (ii) silicon-on-insulator (SOI).  Key to the qubit fabrication process is the use of an anhydrous hydrofluoric vapor process which removes silicon surface oxides without attacking aluminum, and in the case of SOI substrates, selectively removes the lossy buried oxide underneath the qubit region.  For qubits with a transition frequency of approximately $5$~GHz we find qubit lifetimes and coherence times comparable to those attainable on sapphire substrates ($T_{1,\text{Si}}$ = \silife{}, $T_{2,\text{Si}}$ = \sicoh{}; $T_{1,\text{SOI}}$ = \soilife{}, $T_{2,\text{SOI}}$ = \soicoh{}).  This qubit fabrication process in principle permits co-fabrication of silicon photonic and mechanical elements, providing a route towards chip-scale integration of electro-opto-mechanical transducers for quantum networking of superconducting microwave quantum circuits.
\end{abstract}

 % Quantum computation; microwave circuits; superconducting devices
%Use showkeys class option if keyword
\maketitle

%\section{\label{sec:intro} Introduction}

\begin{figure*}
\includegraphics[width=2\columnwidth]{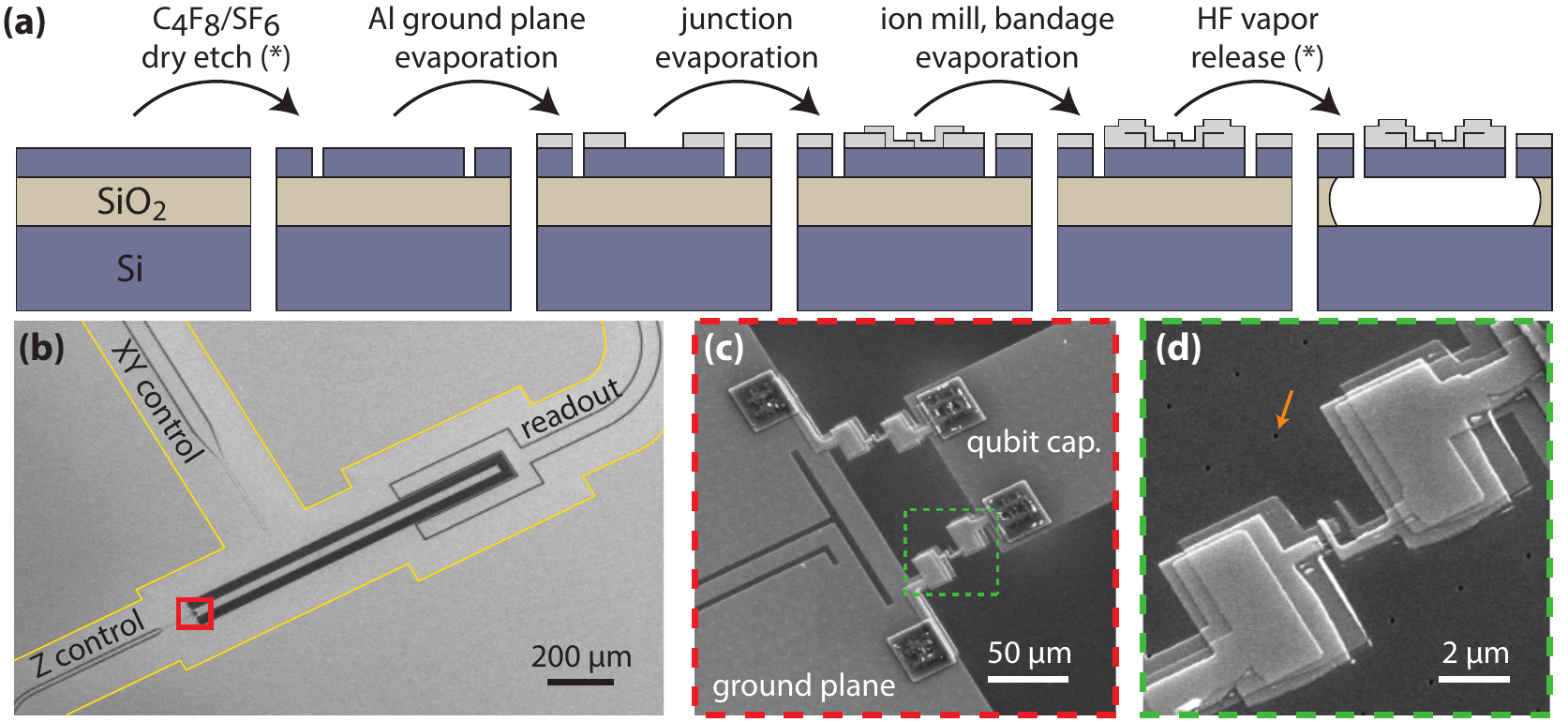}
\caption{\label{fig:fab} Qubit fabrication process and SEM images of the SOI device. (a) Five step fabrication process as detailed in the text. Steps labeled (*) are omitted for the Si qubit process. (b) SEM image of an SOI qubit. The light (dark) gray regions are Al (exposed Si). The yellow outline demarcates the etch front of the HF vapor release, which extends $\approx 100~\mu$m under the ground plane so as to isolate the qubit from the lossy Si-SiO$_2$ interface. The red box denotes the SQUID loop region of the device. (c) Zoom-in image of the SQUID loop, formed by a double angle evaporation process. The green box bounds one junction. ``Bandage" regions described in the main text are visible as darker squares on both the qubit capacitor and the ground plane. (d) Zoom-in of an individual Josephson junction. Each junction has an approximate area of (200~nm)$^2$, corresponding to a zero-bias Josephson Inductance of $L_{J,0} = 22$~nH per junction under the conditions described in the main text. The lattice of tiny dark circles faintly visible here are the etched holes that allow for HF vapor release. An orange arrow points to one such hole.}
\end{figure*}

In recent years, significant developments in experimental quantum information science~\cite{Devoret2013,Houck2012} have been realized using microwave superconducting qubit hardware. These devices, consisting of Josephson junctions (JJs) and linear circuit elements, are typically coupled to high-$Q$ superconducting microwave cavities, which realizes the microwave analog of cavity QED---so-called circuit QED~\cite{Blais2004,Blais2007,Devoret2007}. The advent of the transmon qubit~\cite{Koch2007,Schreier2008,Houck2009} has provided a robust and scalable circuit QED building block. Leveraging small mode volumes and large vacuum coupling rates, circuit QED systems have been put into the regime of strong coupling,~\cite{Wallraff2004,Devoret2007} realized state-of-the-art gate fidelities,~\cite{Barends2014} and utilized to perform quantum error detection and correction.~\cite{Reed2014,Ofek2016} 

Interfacing the circuit QED toolbox with other systems of physical or technological interest -- cavity optomechanical systems, for example~\cite{Safavi-Naeini2011,Andrews2014} -- requires scalable fabrication techniques on compatible materials systems. Much work within the circuit QED community has focused on developing fabrication methods that realize long qubit lifetimes and small dephasing rates~\cite{Quintana2014,Wang2015,McDermott2009}. Owing to sapphire's good microwave properties, this work has primarily utilized the aluminum-on-sapphire (AOS) materials system.  Using the AOS material system two primary approaches have emerged: the so-called planar approach wherein qubits are coupled to on-chip resonators~\cite{Barends2013} and the 3D cavity approach wherein qubits are coupled to 3D box cavities~\cite{Axline2016}. Whereas the former affords higher device densities and more integration, the latter yields longer coherence times.

Here, we demonstrate scalable fabrication techniques for planar silicon-based superconducting circuits~\cite{OConnell2008,Weber2011,Bruno2015} that obtain similar transmon qubit coherence times and gate fidelities as their planar sapphire counterparts~\cite{Barends2013, Barends2014}. We note that similar work has recently been performed with silicon-based qubits in the context of 3D cavities~\cite{Chu2016}.  Additionally, we present the fabrication and characterization of a superconducting qubit on silicon-on-insulator (SOI) with a coherence time which is a factor of $20$ improvement over prior work in this material system~\cite{Patel2013}.  These SOI qubit fabrication methods not only realize high quality qubits, but are also compatible with the integration of other photonic, electronic, and MEMS components on the same SOI substrate.

\begin{figure}
\includegraphics[width=\columnwidth]{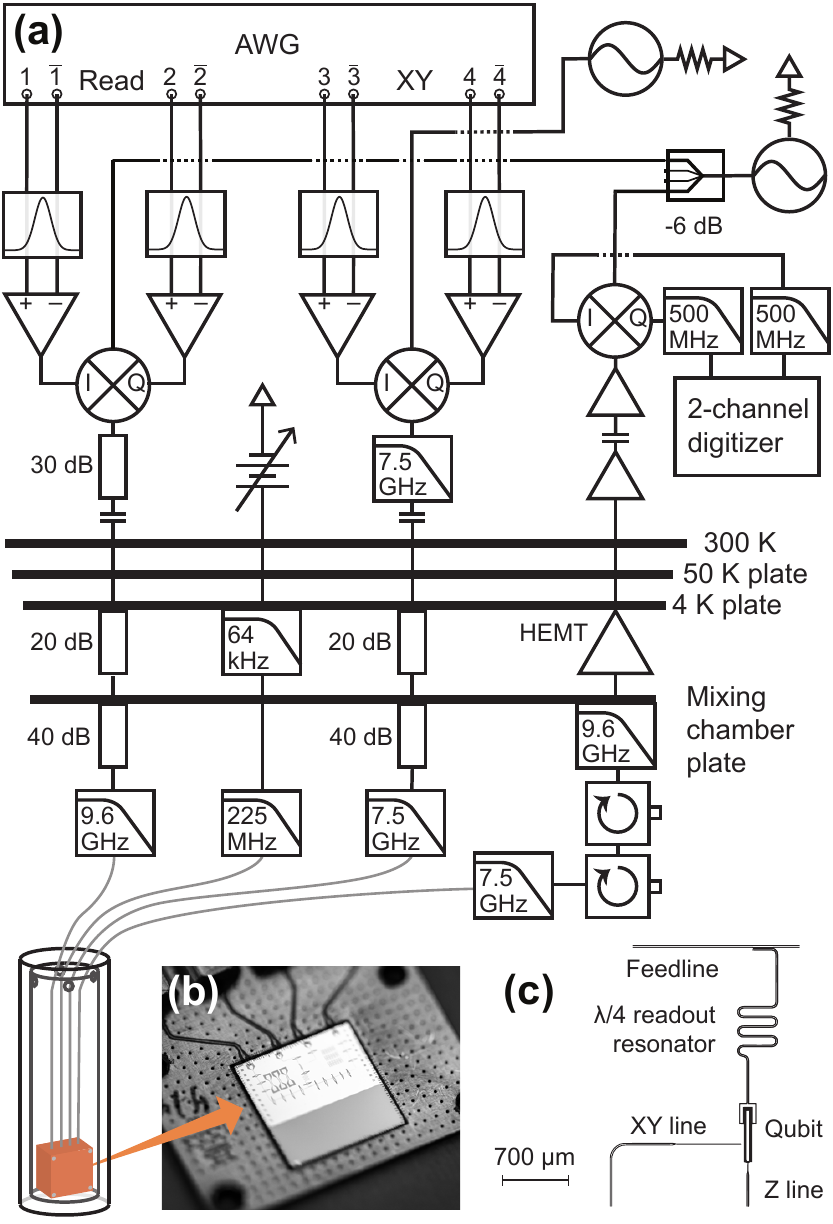}
\caption{\label{fig:meassetup} (a) Time-domain measurement scheme. Near the top, Gaussian filters are indicated by enclosed Gaussian lineshapes (lines $i$ and $\bar{i}$ are filtered individually). CW microwave sources with $Z = $~50~$\Omega$ are indicated by the ac voltage symbols. The microwave source used for readout is followed by a power divider (we use just two ports and terminate others with 50~$\Omega$). Attenuators are indicated by rectangles with labeled power attenuation. Capacitor symbols show inner/outer DC blocks. All low pass filters are reflective except for the 64~kHz filter, which is a dissipative RCR filter (R = 499 $\Omega$, C = 10 nF). (b) Photograph of 1~cm$^2$ chip wire-bonded to PCB. (c) Schematic of the Si and SOI Xmon circuit, including layut of read-out resonator, control lines, and cavity feedline.}
\end{figure}

Our qubit design (pictured in \subfig{\ref{fig:fab}}{b}, and shown schematically in \subfig{\ref{fig:meassetup}}{c}) is based on the Xmon qubit.~\cite{Barends2013} In both our high-resistivity silicon (Si) and SOI devices, a long rectangular capacitor is capacitively coupled to both a readout resonator and an XY-control line; the capacitor is connected to ground through a SQUID loop (\subfig{\ref{fig:fab}}{c}) that is inductively coupled to a DC control line, which allows for frequency tuning of the qubit.~\cite{Barends2013} Our readout resonator, consisting of a $\lambda/4$ coplanar waveguide resonator, is inductively coupled to a transmission line, which allows for dispersive readout of the qubit.~\cite{Barends2013} We realize (Si, SOI) as-measured parameters of: $f_q = \omega_q/2\pi = (4.962,5.652)$~GHz, %4.9618,5.6520
$\eta/2\pi = (-260,-300)$~MHz, %259$\pm$2 MHz and 296$\pm$8 MHz
$\omega_r/2\pi = (6.868,7.143)$~GHz, %6.8675,7.1426
and $\chi/2\pi = (1.2,3.5)$~MHz, where $\omega_q=\omega_{10}$ is the qubit transition frequency, $\eta = (\omega_{21} - \omega_{10})$ is the anharmonicity, $\omega_r$ is the readout resonator frequency, and $2\chi = \omega_{r,|0\rangle} - \omega_{r,|1\rangle}$ is the dispersive shift. These measured values imply a Josephson energy $E_J/h = (13.1,14.8)$~GHz in the transmon limit ($E_J \gg E_C$) where $\hbar \omega_q \approx \sqrt{8E_JE_C}-E_C$ and the charging energy $E_C \approx -\hbar\eta$, as well as a vacuum qubit-resonator coupling rate $g/2\pi = (135,177)$~MHz where $g \approx \sqrt{-\Delta \chi (1+\Delta/\eta)}$ and $\Delta = \omega_q - \omega_r$. %The qubit-resonator detunings are $\Delta/2\pi = (1.906,1.491)$~GHz. 
Our readout resonators have intrinsic and extrinsic coupling $Q$s of $Q_i = (5.8,45.8)\times 10^3$ and $Q_e = (12.9,6.1)\times 10^3$, respectively, measured at single-digit intracavity photon numbers.  These values are close to the designed and expected values, except for the intrinsic $Q_i$ of the read-out resonator on Si.  This value is more then two-orders of magnitude smaller than expected from previous resonator-only tests we have performed on Si.  Evidence of frequency jitter in the read-out resonator of this sample was observed, which may explain an under-estimate of the $Q_i$ from the swept frequency measurement used here.    
% E_{J,Si} = (4.962e9 + 260e6)^2/8/260e6 => E_J/E_c ~ 50.4
% E_{J,SOI} = (5.652e9 + 300e6)^2/8/300e6 => E_J/E_c ~ 49.3
% g=sqrt(-\Delta*\chi*(1+\Delta/\eta))
% for Si: sqrt(-0.00115*(4.962-6.868)*(1+(4.962-6.868)/-0.259))
% for SOI: sqrt(-0.0035*(5.6520-7.1426)*(1+(5.6520-7.1426)/-0.296))
% see Sank's thesis, note that \omega_r > \omega_q => \Delta <0
% note also that anharmonicity should be negative for a transmon

Our fabrication process is a multi-layer process pictured in \subfig{\ref{fig:fab}}{a}. We begin with a 10~mm $\times$ 10~mm chip of either Si [Float zone (FZ) grown, $525$~$\mu$m thickness, $>10$~kOhm-cm resistivity] or SOI [Si device layer: FZ grown, $220$~nm, $\gtrsim 3$~kOhm-cm; BOX layer: $3$~$\mu$m, silicon dioxide; Si handle: Czochralski grown, $750$~$\mu$m, $\gtrsim 5$~kOhm-cm]. We then perform the following main fabrication steps (from left to right in \subfig{\ref{fig:fab}}{a}): (i) C$_4$F$_8$/SF$_6$ inductively coupled plasma reactive ion etch (ICP-RIE) of $50$~nm radius holes through the device layer to allow for release in step (v) below; (ii) electron beam evaporation of $120$~nm Al at $1$~nm/s to define a ground plane, the qubit capacitor, and the readout resonator; (iii) double-angle electron beam evaporation of $60$~nm and $120$~nm of Al at $1$~nm/s with an intervening $20$~minute oxidation at $5$~mbar and subsequent $2$~minute oxidation at $10$~mbar to forms the JJs; (iv) $5$ minute argon ion mill and $140$~nm Al evaporation to form a ``bandage" layer that electrically contacts the Al layers defined in step (ii) and (iii); (v) HF vapor etch of the underlying BOX layer.

After steps (ii-iv), a liftoff process was performed in n-methyl-2-pyrrolidone at $80$~$^{\circ}$C for two hours. Steps (i) and (v) are omitted for Si devices as they do not require a release. In (i-iv), we use electron beam lithography to pattern our resist. The above process is similar to that described elsewhere~\cite{Dieterle2016,Megrant2016} and, for SOI samples, yields a device layer that is partially suspended above the handle wafer. As highlighted by the yellow boundary line in the scanning electron microscope image of \subfig{\ref{fig:fab}}{b}, we etch $100$~$\mu$m into the BOX layer such that the circuit is far from the lossy Si/SiO$_2$ interface.~\cite{Dieterle2016}

We characterize each qubit in a $^{3}$He/$^{4}$He dry dilution refrigerator with base temperature of $T_f \sim 7$~mK using frequency-domain and time-domain spectroscopy. We begin with frequency-domain characterization and measure transmission ($S_{21}$) through a coplanar waveguide feedline using a two-port vector network analyzer (VNA). The Z control line is used to carry a small current which produces an external flux bias, $\Phi_{\rm ext}$, in the SQUID loop of the qubit, thereby tuning the qubit transition frequency, $f_q$. %The qubit transition is not directly observable, but for weak probe powers we can observe the readout resonance shifting periodically in external flux bias $\Phi_{\rm ext}$ applied to the qubit, owing to $\chi$ depending on $\Delta$.
For a given $\Phi_{\rm ext}$, we identify $f_q$ and transitions to higher levels (from which we extract $\eta$) by sweeping a continuous-wave (CW)  microwave tone applied to the XY drive line and monitoring the resonator response.~\cite{FinkThesis}

\begin{figure}
\includegraphics[width=\columnwidth]{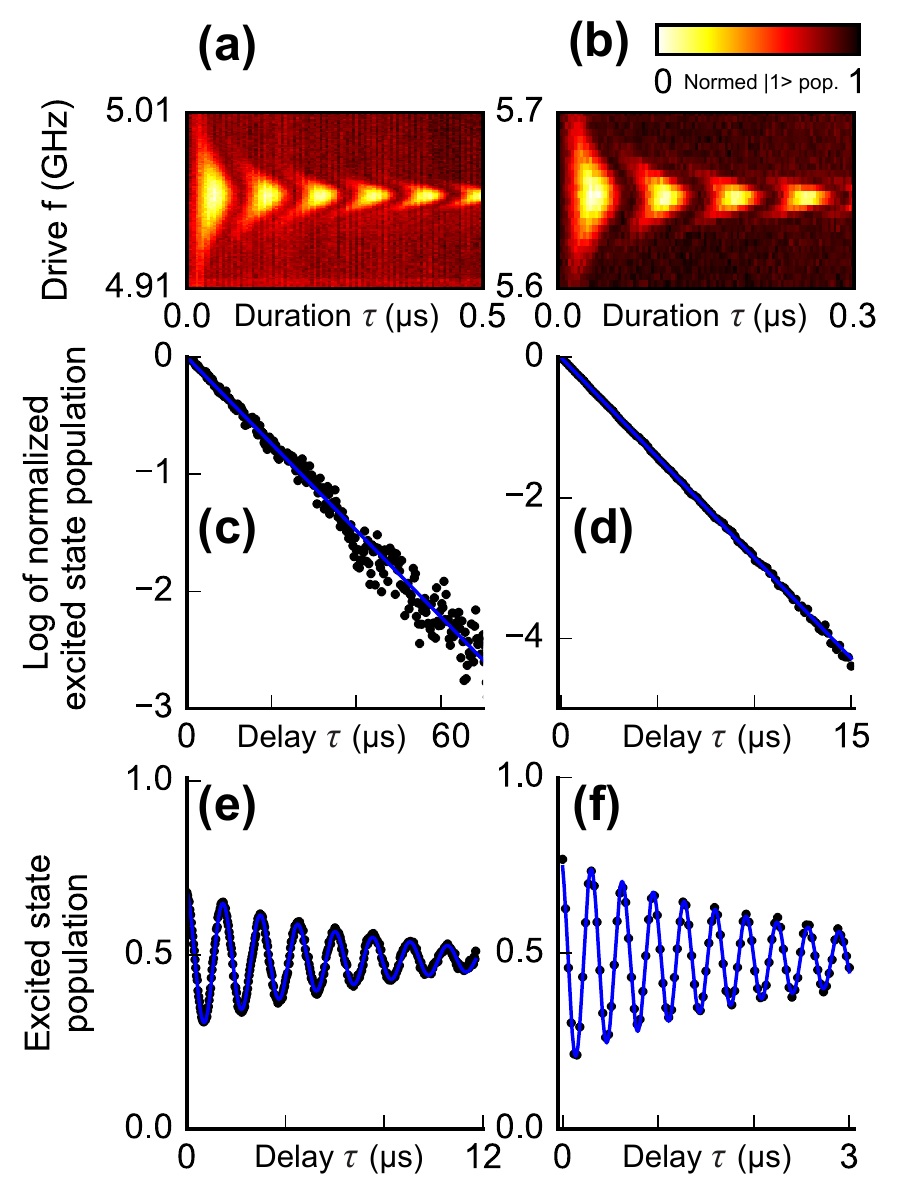}
\caption{\label{fig:t1t2} Qubit characterization (left column: Si; right column: SOI). (a,b) Excited state population (normalized to the unit interval) as a function of XY drive frequency and pulse duration $\tau$ exhibits a chevron pattern typical of a qubit undergoing Rabi oscillations. (c,d) Natural log of the excited state population, normalized to the unit interval, shows exponential decay as a function of waiting time $\tau$ with lifetimes $T_{1,\rm Si} =$~\silife{} and $T_{1,\rm SOI} = $~\soilife{}. (e,f) By applying two off-resonance $\pi/2$ pulses with a variable intervening delay $\tau$, the excited state population shows Ramsey oscillations (points are data, blue trace is fit). The decay of the envelope yields coherence times $T_{2,\rm Si} =$~\sicoh{} and $T_{2,\rm SOI} =$~\soicoh{}. In all cases, we use a rectangle-windowed readout pulse with 500~ns duration, and in most cases we use a 30~ns X$_\pi$ and X$_{\pi/2}$ pulse ($45$~ns in (c)).}
\end{figure}

Having identified device parameters, we switch over to time-domain characterization, using the measurement setup summarized in \fig{\ref{fig:meassetup}} (for details see App.~\ref{App:A}).  We characterize each qubit using dispersive readout~\cite{Schuster2005} (\fig{\ref{fig:t1t2}}) with $\Phi_{\rm ext}$ set so that the qubit is at a first-order flux-insensitive point.~\cite{Koch2007,Barends2013} In this projective readout scheme, a sample from one of two distributions in the I-Q plane is measured depending on whether the qubit is projected into the ground or excited state. We make a binary discrimination for each single measurement of an arbitrary pulse sequence ($1$ for excited qubit state $|1\rangle$, or $0$ for ground qubit state $|0\rangle$), and average at least $10^4$ such values %sometimes up to 500000
to determine excited state population. The sub-unity visibility (e.g., see \subfig{\ref{fig:t1t2}}{e,f}) primarily reflects imperfect readout fidelity for both Si and SOI devices.~\cite{readoutdetail}

\begin{figure}
\includegraphics[width=\columnwidth]{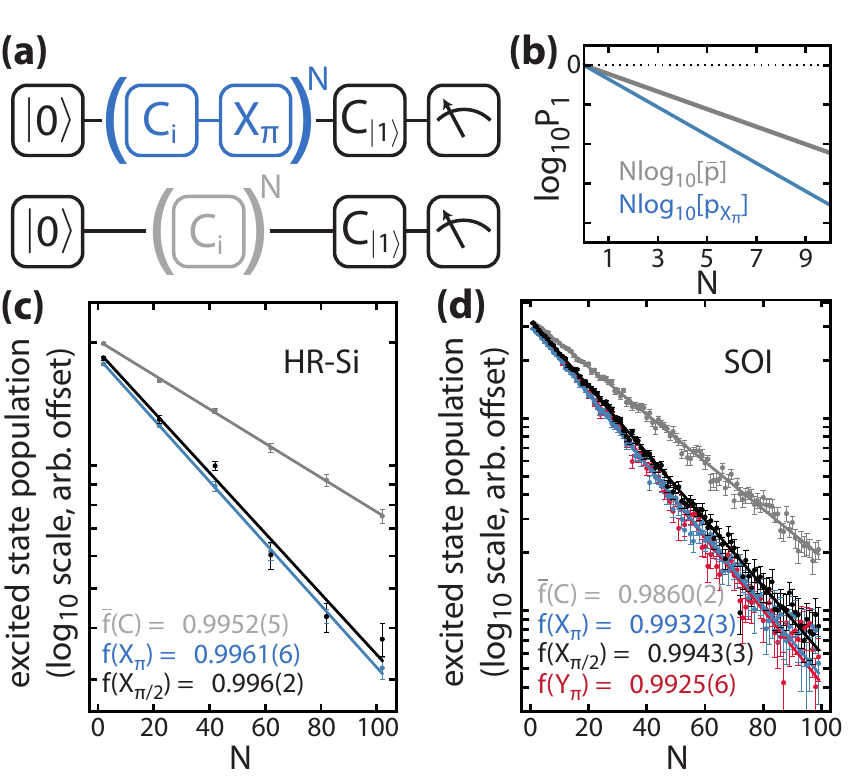}
\caption{\label{fig:RB} Randomized benchmarking. (a) A schematic of Clifford Group randomized benchmarking, described in detail in App.~\ref{App:B}. (b) A plot of the excited state probability as a function of $N$ reveals the gate fidelity through the slope of the resultant line on a semilog plot and the relations described in App.~\ref{App:B}. The limit of perfect fidelity is shown as a dashed line. (c, d) Plots of the gate fidelity (with an arbitrary offset given by the readout fidelity) as a function of $N$ for both Si (c) and SOI (d) qubits. Error bars in the plots represent 1 standard error in the measurements averaged over $(40,50)$ random Clifford sequences on (Si,SOI). Error bars in the gate fidelities represent 1 standard deviation of $f$ due to the statistical uncertainty of the parameter $p$ in the exponential fit.}
\end{figure}

%The circulators are RADC-4-8-Cryo-0.02-4K-S23-1WR-b (Raditek, San Jose, CA USA). 
%SS-SS .085”, NbTi-NbTi 0.085"

To characterize our gate fidelities, we utilized Clifford group randomized benchmarking,~\cite{Chow2009,Magesan2012,Barends2014} shown schematically in \subfig{\ref{fig:RB}}{a,b}.  For the Si sample \subfig{\ref{fig:RB}}{c}, we measured two gates ($X_{\pi/2}$ and $X_\pi$) while for SOI (\subfig{\ref{fig:RB}}{d}), we measured three ($X_{\pi}$, $X_{\pi/2}$, and $Y_\pi$). We realize average gate fidelities of $\bar{f}(C) = 0.9952(5)$ on Si and $0.9860(2)$ on SOI as well as individual gate fidelities of $> 0.992$ for all measured gates on Si and SOI. It should be noted that these gates have not yet been optimized to avoid phase errors or leakage outside the computational basis~\cite{Motzoi2009}.

Our measurement of $T_{1,\text{Si}} = $~\silife{} for our Si qubit is noteworthy, especially given the simplicity of our design and fabrication process. We can estimate the Purcell-limited $T_1$ by the simplistic single-mode estimate $(\Delta/g)^2/ \kappa_r$, where $\kappa_r = \omega_r/Q$ and $1/Q = 1/Q_i + 1/Q_e$. This yields 18.5~$\mu$s (8.5~$\mu$s) for the Si (SOI) qubit of this work, implying that: (i) the estimate is inaccurate, since we measure a larger-than-estimated $T_1$ for Si (a more conservative estimate which assumes we are indeed under-estimating $Q_i$ due to frequency jitter and takes $Q \approx Q_{e}$, yields a Purcell-limited $T_1$ of 57~$\mu$s), and (ii) incorporating an on-chip Purcell filter may well improve our qubit lifetimes.~\cite{Reed2010,Jeffrey2014} Also, regarding the measured $T_2$ values, since obtaining these measurements we have identified and resolved some grounding issues in our measurement setup that likely contributed to excess flux noise coming from $60$~Hz currents on our flux bias line. We anticipate that these improvements may even be important at the first-order flux insensitive point.

% Si Purcell limit = 1/((g/\Delta)^2*\kappa) where \kappa = w_r/Q
%1/((2*pi*6.868e9/(1/5800+1/12900)^-1*(135e6/(6.868e9-4.962e9))^2))
% = 18.5 us (!!)
%SOI Purcell limit = 1/((g/\Delta)^2*\kappa) where \kappa = w_r/Q
%1/((2*pi*7.143e9/(1/6100+1/45800)^-1*(177e6/(7.143e9-5.652e9))^2))
% = 8.5 us (!!)

In terms of the impact of the SOI device layer properties or various fabrication steps on the resulting qubit decoherence times, further systematic studies are required. In particular, the importance of using the vapor HF etch to remove native oxides and (temporarily) passivate the Si surface before every evaporation step of aluminum on the Si layer (including right before the double angle evaporation used to form the JJs), needs to be clarified further.  Also, any residual effects of the underlying BOX layer needs to be ruled out through systematic studies of qubit coherence versus undercut extent, in conjunction with 3D numerical modeling to determine more optimized qubit and membrane geometries.  Even while the precise physical and materials limitations of our system are unclear, current coherence times are sufficient for many quantum simulation and quantum optics experiments. Meanwhile, our realization of a highly coherent SOI qubit represents an essential building block for hybrid electro-opto-mechanical systems on SOI. Already, electromechanical and optomechanical coherent transduction bandwidths exceed the bandwidth of our qubit by a factor of two~\cite{Andrews2014,Dieterle2016,Witmer2016}, a prerequisite for high-fidelity, bi-directional microwave-to-optical quantum state transduction---an interesting and challenging research program in its own right, with many potential realizations.

Overall, our fabrication and measurements of planar qubits on silicon and SOI represent a modest but important technical stepping stone on the path to a variety of potential quantum information and quantum science goals. Taken together with complementary advances in the fields of cavity opto- and electro-mechanics~\cite{Aspelmeyer2014,Andrews2014}, and in the context of competing systems~\cite{Chu2016,Barends2013}, we are optimistic about the potential for hybrid quantum systems and circuit QED on silicon and silicon-on-insulator platforms.

\section*{Acknowledgments}

We gratefully acknowledge the Martinis Group (UCSB/Google) for their amplifier and filter designs as well as Dan Vestyck at SPTS for his support of our uEtch HF vapor tool. This work was supported by the AFOSR MURI Quantum Photonic Matter (grant 16RT0696), the AFOSR MURI Wiring Quantum Networks with Mechanical Transducers (grant FA9550-15-1-0015), the Institute for Quantum Information and Matter, an NSF Physics Frontiers Center (grant PHY-1125565) with support of the Gordon and Betty Moore Foundation, and the Kavli Nanoscience Institute at Caltech.  A.J.K. acknowledges an IQIM Postdoctoral Fellowship.

%\bibliography{./refs}

\providecommand{\noopsort}[1]{}\providecommand{\singleletter}[1]{#1}%
%

%\onecolumngrid
\appendix

\section{Measurement setup}
\label{App:A}
A Tektronix AWG5014C arbitrary waveform generator (AWG) generates shaped in-phase (I) and quadrature (Q) pulses at IF~$= 100$~MHz for both qubit readout and XY drive. Each output of the AWG passes through its own home-made dissipative Gaussian filter with 320~MHz cutoff. The waveforms are each amplified with a home-made differential amplifier and passed to the I and Q ports of IQ mixers (Marki IQ-0307MXP for the XY drive, IQ-0409MXP for readout). Carrier tones are supplied by CW microwave sources (Rohde~\&~Schwarz~SMB100A) to the local oscillator (LO) ports of the mixers. As a result, the readout and XY pulses are single-sideband-upconverted to microwave frequencies. We attenuate and filter these signals at several temperature stages of a cryogen-free dilution refrigerator.

Flux biasing is provided by a programmable DC source (Yokogawa GS200) which is filtered at 4~K (Therma-uD-25G from Aivon Oy, Helsinki, Finland) and again at the mixing chamber plate with a reflective microwave filter (Minicircuits).

The DC and AC signals reach the device, which is mounted on a gold-plated PCB inside a copper box inside two concentric magnetic shields (Magnetic Shields Ltd., Staplehurst, UK) consisting of 1.5~mm thick Cryophy material heat-treated to MSL1154-HTC specification. The inner shield is 51mm ID by 168mm high and the outer shield is 67mm ID by 185mm high. The copper box and magnetic shields are mounted to a copper coldfinger attached to the mixing chamber plate. A shield on the mixing chamber is painted in an infrared-absorbing carbon/silica/epoxy mixture to minimize quasiparticle generation in the aluminum.~\cite{Barends2011,Corcoles2011}

The output is protected from room-temperature noise by two circulators (Raditek RADC-4-8-Cryo-0.02-4K-S23-1WR-b) and at 4~K, a HEMT (Low Noise Factory LNF-LNC4\_8C) amplifies by 42~dB, with 68~dB further amplification at room temperature. We used two room-temperature power amplifiers, a Miteq AFS42-00101200-22-10P-42 with 50~dB of gain and a home-made amplifier with 18~dB of gain designed by the Martinis group. The readout signal is then downconverted and the resulting I and Q are simultaneously digitized using a 1GS/s 2-channel PCIe digitizer (AlazarTech ATS9870). In software, the I and Q are mixed with 100~MHz tones to yield a single point in the I-Q plane for a single readout pulse.

The semirigid coaxial cable in our fridge is stainless-stainless .085'' above the 4~K plate and NbTi-NbTi .085'' below. 

\section{Randomized benchmarking}
\label{App:B}

In Clifford group randomized benchmarking protocols,~\cite{Magesan2012,Barends2014,Chow2009} a qubit initialized in its ground state has $2N$ gates performed on it---$N$ random gates from the Clifford group (labeled $C_i$) interleaved with $N$ of the same Clifford gate (e.g., $X_\pi$). After application of all $2N$ gates, we perform the Clifford gate that puts the qubit in its excited state (labeled $C_{|1\rangle}$) and read out the qubit state. The probability of being in the excited state as a function of $N$, is then compared against the same procedure, but without $N$ of the same interleaved gate. This procedure yields two exponentials of the form $c_1+c_2p^N$, the latter with depolarizing parameter $\bar{p}$ and the former with depolarizing parameter $p_G$. These two parameters are related to the average Clifford group gate fidelity, $\bar{f}(C)$, and gate fidelity of interest, $f(G)$, through $\bar{f}(C)=1-\frac{1-\bar{p}}{2}$ and $f(G)=1-\frac{1-p_G/\bar{p}}{2}$.~\cite{Magesan2012} For the data presented in Fig. 4, we performed 20000 measurements on 40 (50) random Clifford group sequences on Si (SOI).

\end{document}